\documentclass[a4paper,11pt]{article}
\usepackage{pos}

\title{LHC sensitivity to $Z^\prime / W^\prime$ states in composite Higgs models}
\ShortTitle{LHC sensitivity to $Z^\prime / W^\prime$ states in composite Higgs models}

\author*[a]{J.~Fiaschi}
\author[b]{F.~Giuli} 
\author[b,c,d]{F.~Hautmann}
\author[e,f]{S.~Moretti}

\affiliation[a]{Department of Mathematical Sciences, University of Liverpool, Liverpool L69 3BX}
\affiliation[b]{CERN, European Organization for Nuclear Research, CH 1211 Meyrin}
\affiliation[c]{Elementaire Deeltjes Fysica, Universiteit Antwerpen, B 2020 Antwerpen}
\affiliation[d]{Theoretical Physics Department, University of Oxford, Oxford OX1 3PU}
\affiliation[e]{School of Physics and Astronomy, University of Southampton, Southampton SO17 1BJ}
\affiliation[f]{Department of Physics and Astronomy, Uppsala University, SE-751 20 Uppsala}

\emailAdd{Juri.Fiaschi@liverpool.ac.uk}
\emailAdd{francesco.giuli@cern.ch}
\emailAdd{hautmann@thphys.ox.ac.uk}
\emailAdd{S.Moretti@soton.ac.uk}
\emailAdd{stefano.moretti@physics.uu.se}

\abstract{Using the 4-Dimensional Composite Higgs Model (4DCHM) realization of the minimal composite Higgs scenario, we discuss the Large Hadron Collider (LHC) sensitivity to new physics signals from multiple $Z^\prime$ and $W^\prime$ broad resonances. We illustrate the role of systematic uncertainties due to QCD effects encoded in parton distribution functions for experimental searches in leptonic channels. 
We show that, by  reducing this systematics through the combination of high-precision measurements of Standard Model (SM) lepton-charge and forward-backward asymmetries near the SM vector-boson peak, the sensitivity to the new physics signals can be greatly enhanced.} 

\FullConference{%
  41st International Conference on High Energy physics - ICHEP2022\\
  6-13 July, 2022\\
  Bologna, Italy
}

\begin{document}

\vspace*{-0.6cm} 
\hspace*{11.0cm} CERN-TH-2022-200\\  \hspace*{13.4cm} LTH 1326

\maketitle

Experimental studies of di-electron, di-muon and charged leptons missing transverse energy channels constitute classic methods to search for Beyond-the-Standard-Model (BSM) $Z^\prime / W^\prime$ gauge bosons at the Large Hadron Collider (LHC)~\cite{ATLAS:2019erb,ATLAS:2019lsy,CMS:2018ipm,CMS:2021ctt}, and will be further pursued at the High-Luminosity LHC (HL-LHC)~\cite{CidVidal:2018eel}. 

In the case of narrow vector resonances, experimental analyses rely on resonant mass searches based on the Breit-Wigner (BW) lineshape. In the case of vector resonances with large width, on the other hand, this is likely not sufficient, as instead of an easily observable narrow BW lineshape one has a broad shoulder spreading over the Standard Model (SM) background.
Alternative experimental approaches then have to be applied, which are generally much more dependent than resonant mass searches on the modeling of the production process, for both signal and background~\cite{Accomando:2019ahs}. 
In this case, one of the main sources of systematic uncertainties affecting the experimental sensitivity to new physics is given by initial-state Quantum Chromodynamics (QCD) effects~\cite{Fiaschi:2022wgl,Ball:2022qtp}. 

Additional features influencing the experimental search strategy arise in scenarios with multiple broad resonances. 
In these scenarios, interference effects between the BSM resonances themselves and between the BSM and SM states can give rise to a statistically significant depletion of events below the BW peak in the invariant and transverse mass distributions, leading to the appearance of a pronounced {\em dip}~\cite{Accomando:2019ahs,Fiaschi:2021okg}. 
The significance of the depletion of events may be defined in a manner similar to that for the excess of events of the peak, so that BSM exclusion and discovery limits may be extracted from the analysis of the mass spectra for either the peak or the dip. 

Gauge sectors of strongly-coupled models of electroweak symmetry breaking~\cite{Panico:2015jxa} based on composite Nambu-Goldstone 
Higgs~\cite{Kaplan:1983fs,Kaplan:1983sm} provide examples with multiple $Z^\prime $ and $ W^\prime$ resonances, featuring large width and interference effects. 
In this article, based on Ref.~\cite{Fiaschi:2021sin}, we consider the 4-Dimensional Composite Higgs Model (4DCHM) realization~\cite{DeCurtis:2011yx} of the minimal composite Higgs scenario~\cite{Agashe:2004rs}. The parameter space of the model can be 
described in terms of the compositeness scale $f$ and the coupling $g_\rho$ of the new resonances, with the resonances mass scale 
being of order $M \sim f g_\rho$~\cite{Panico:2015jxa,Giudice:2007fh}. 
We select two 4DCHM benchmarks, each characterized by specific values of $f$, $g_\rho$ and the resonance masses. We analyze the role of initial-state QCD effects, encoded in the evolution of parton distribution functions (PDFs), in the determination of exclusion and discovery limits in the parameter space of these two 4DCHM benchmarks, focusing on leptonic channels. (See e.g.~\cite{Liu:2019bua} for a study comparing leptonic with heavy-quark channels in composite Higgs models.)

The main observation underlying the analysis~\cite{Fiaschi:2021sin} is that PDF systematic uncertainties can be improved by exploiting the experimental information on the SM vector bosons polarization in the mass range near the SM vector bosons peaks. 
This has been explored in the case of the longitudinal~\cite{Amoroso:2020fjw} and transverse~\cite{Fiaschi:2021okg,Accomando:2019vqt} vector-boson polarizations via ``profiling'' studies using the \texttt{xFitter}~\cite{xFitter:2022zjb,Alekhin:2014irh} QCD analysis platform. In particular, ``improved PDFs'' are obtained in Ref.~\cite{Fiaschi:2021okg} by combining high-statistics precision measurements of lepton-charge $A_W$ and forward-backward $A_{\rm{FB}}$~\cite{Accomando:2019vqt,Abdolmaleki:2019ubu,Accomando:2018nig,Accomando:2017scx} asymmetries, associated with the difference between the left-handed and right-handed vector boson polarization fractions. 
One can then investigate the impact of improved PDFs on the experimental sensitivity to the BSM searches in the mass region defined by the current LHC exclusion limits~\cite{ATLAS:2019erb,ATLAS:2019lsy,CMS:2018ipm,CMS:2021ctt}, in which evidence for $Z^\prime / W^\prime$ states with large widths could first be observed. 

\begin{table}
\begin{center}
\begin{tabular}{|c|c|c|c|}
\hline
\multicolumn{4}{|c|}{Benchmark A}\\
\hline
inf [TeV] & sup [TeV] & $\sigma_{\rm SM}$ [fb] & $\sigma_{\rm SM+BSM}$ [fb] \\
2.06 & 4.99 & 1.69 $\cdot$ 10$^{-1}$ & 1.42 $\cdot$ 10$^{-1}$ \\
\hline
$\Delta_{\rm PDF}$ base [fb] & $\Delta_{\rm PDF}$ profiled [fb] & $\alpha$ (base) & $\alpha$ (profiled)\\
9.5 $\cdot$ 10$^{-3}$ & 4.6 $\cdot$ 10$^{-3}$ & 3.34 & 4.82 \\
\hline
\end{tabular}
\begin{tabular}{|c|c|c|c|}
\hline
\multicolumn{4}{|c|}{Benchmark B}\\
\hline
inf [TeV] & sup [TeV] & $\sigma_{\rm SM}$ [fb] & $\sigma_{\rm SM+BSM}$ [fb] \\
1.36 & 3.36 & 1.53 & 1.45 \\
\hline
$\Delta_{\rm PDF}$ base [fb] & $\Delta_{\rm PDF}$ profiled [fb] & $\alpha$ (base) & $\alpha$ (profiled)\\
6.8 $\cdot$ 10$^{-2}$ & 3.1 $\cdot$ 10$^{-2}$ & 1.53 & 2.91 \\
\hline
\end{tabular}
\end{center}
\caption{The table reports, for the benchmarks 
A and B \cite{Fiaschi:2021sin}: (i) integration intervals in invariant mass for the dip region; (ii) resulting cross sections for the SM background and the complete 4DCHM; (iii) associated PDF uncertainty with the baseline CT18NNLO set~\cite{Hou:2019efy} and with the 
profiled set~\cite{Fiaschi:2021okg} using the 
$A_{\rm FB} + A_W$ combination; (iv) significances $\alpha$ employing the two PDF errors for an integrated luminosity of 3000 fb$^{-1}$.}
\label{tab:AB_dip_neutral}
\end{table}

\begin{figure}
\begin{center}
\includegraphics[width=0.48\textwidth]{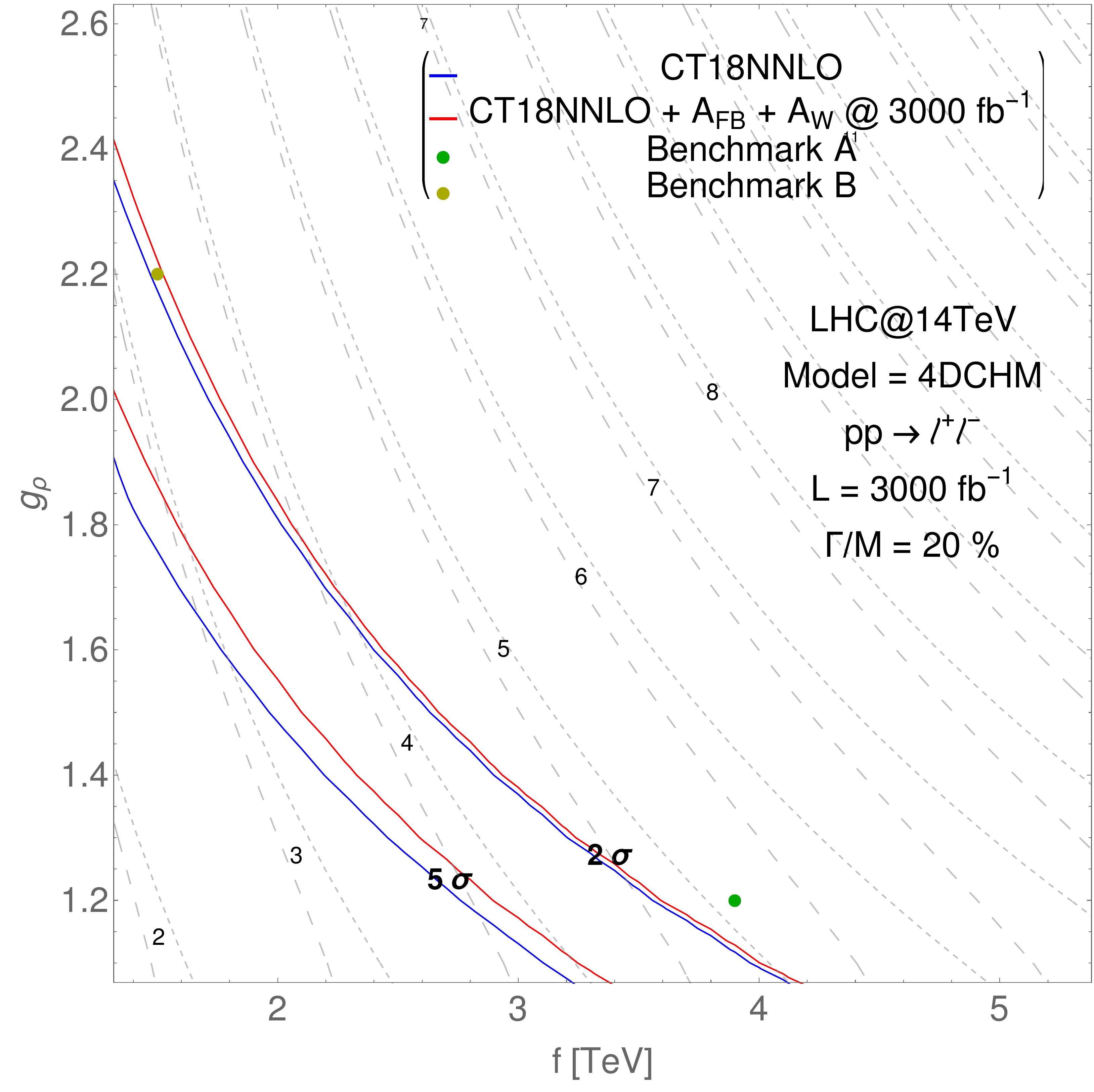}
\includegraphics[width=0.48\textwidth]{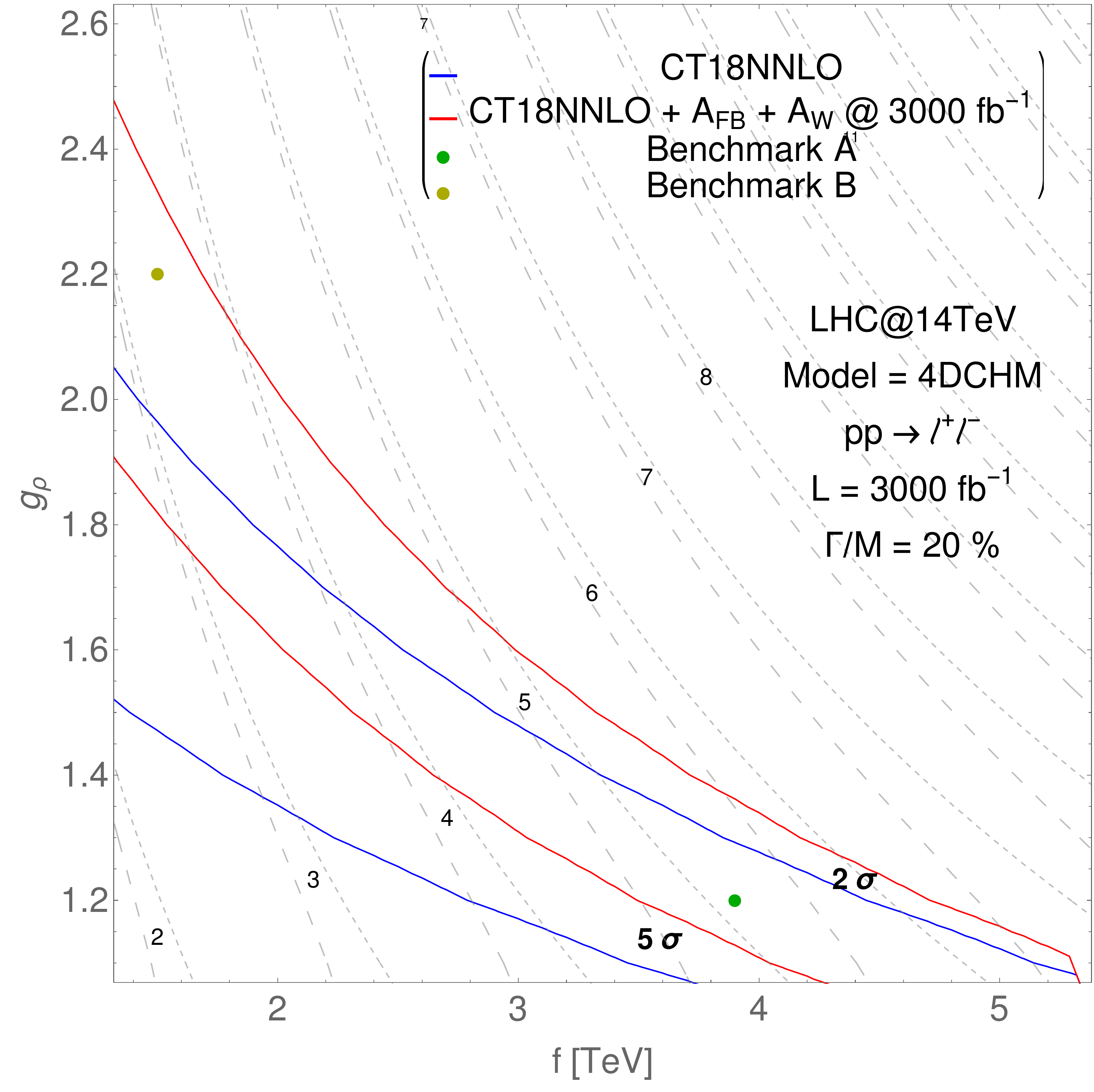}
\end{center}
\caption{Exclusion and discovery limits at 3000 fb$^{-1}$ for the peak (left) and for the dip (right) for $Z^{\prime}$ resonances with $\Gamma / M$ = 20\%~\cite{Fiaschi:2021sin}.
The short (long) dashed contours give the BSM boson mass $M_{Z_2}$ ($M_{Z_3} \simeq M_{W_2}$) in TeV.}
\label{fig:Contour_20_3000}
\end{figure}

Fig.~\ref{fig:Contour_20_3000} shows an example of this analysis for the two 4DCHM benchmarks defined in~\cite{Fiaschi:2021sin}, denoted by A and B.
In Fig.~\ref{fig:Contour_20_3000} limits from the neutral current di-lepton channel are shown on the model parameter space, described in terms of $f$ and $g_\rho$, for the HL-LHC (center-of-mass energy of 14 TeV, integrated luminosity of 3000 fb$^{-1}$). 
In the background, contour plots are given for the masses of the new gauge bosons. 
The blue and red curves are obtained respectively with the baseline CT18NNLO PDF set~\cite{Hou:2019efy} and with the profiled PDFs~\cite{Fiaschi:2021okg} using the $A_{\rm{FB}}$ and $A_W$ combination with 3000 fb$^{-1}$ of integrated luminosity.
The left panel in Fig.~\ref{fig:Contour_20_3000} is for the peak analysis, the right panel is for the dip analysis.
In this set-up, the peak of benchmark A would still be below the experimental sensitivity, while the peak of benchmark B, if the improved PDFs are employed, would be right below the 2$\sigma$ exclusion.
When exploiting the depletion of events in the dip below the peak, the sensitivity on the model increases remarkably.
The PDF improvement has a large impact especially in the region of small $f$ and large $g_\rho$. Here the sensitivity on the dip can overtake the LHC reach in the peak region once the profiled PDFs are employed. 
Taking into account the reduction of PDF uncertainty, benchmark A would be now at the edge of the 5$\sigma$ discovery, while the sensitivity on benchmark B would almost reach 3$\sigma$.

Tab.~\ref{tab:AB_dip_neutral} \cite{Fiaschi:2021sin} gives numerical results for the dip analysis, in the case of the two benchmarks A and B. We see that the improvement in significance due to the profiled PDFs is sizeable, as significances grow by an amount from 40\% to 90\% (i.e., by a factor of about 10 in comparison) for the benchmarks A and B, respectively. 
A more modest improvement in significance is found \cite{Fiaschi:2021sin} in the case of the peak. 

Analogous studies are performed also for the charged current sector in~ \cite{Fiaschi:2021sin}. In cases in which both neutral and charged new states appear and are correlated by theory, the reduction of systematic PDF error can usefully be applied to combined $W^\prime$ and $Z^\prime$ searches. For example, in the 4DCHM a direct $W^\prime$ exclusion (or indeed discovery) achieved in the charged-current channel can be used indirectly to probe the existence of a $Z^\prime$ better than this can be done with direct searches in the neutral-current channel.
The analysis of the dip often provides more stringent limits than the signal coming from the peak, with the reduction of systematic PDF errors playing a crucial role in this conclusion.

\vskip 0.3 cm 

\section*{Acknowledgments}
\noindent
We thank the ICHEP2022 organizers and convenors for the conference and the invitation.
The work of J.~Fiaschi has been supported by STFC under the Consolidated Grant ST/T000988/1.
S.~Moretti is supported in part through the NExT Institute and acknowledges funding from the STFC Consolidated Grant ST/L000296/1.

\end{document}